# Effect of Share Capital on Financial Growth of Non-Financial Firms Listed at the Nairobi Securities Exchange

*David Haritone Shikumo[1]*

*Abstract*
***Purpose:*** *A significant number of the non-financial firms listed at the Nairobi Securities Exchange (NSE) have been experiencing declining financial performance which deters investors from investing in such firms. The lenders are also not willing to lend to such firms. As such, the firms struggle to raise funds for their operations. Prudent financing decisions can lead to financial growth of the firm. The purpose of this study is to assess the effect of Share capital on financial growth of Non-financial firms listed at the Nairobi Securities Exchange. Financial firms were excluded because of their specific sector characteristics and stringent regulatory framework. The study is guided by Market Timing Theory and Theory of Growth of the Firm.*
***Methodology:*** *Explanatory research design was adopted. The target population of the study comprised of 45 non-financial firms listed at NSE for a period of ten years from 2008 to 2017. The study conducted both descriptive statistics analysis and panel data analysis.*
***Findings****: The result indicates that, share capital explains 32.73% and 11.62% of variations in financial growth as measure by growth in earnings per share and growth in market capitalization respectively. Share capital positively and significantly influences financial growth as measured by both growth in earnings per share and growth in market capitalization.*
***Implication****s: The study recommends for the Non- financial firms to utilize equity financing as a way of raising capital for major expansions, asset growth or acquisitions which may require heavy funding. In this way, firms will be assured of improved performance as well as high financial growth. The study also recommends for substantial firm financing through equity.*
***Value****: Equity financing is important to any firm, if the proceeds are used to invest in projects which eventually bring growth to the firm.*

---

[1] PhD Finance Student, Jomo Kenyatta University of Agriculture and Technology, Email: dashikumo@gmail.com

## 1.1 Introduction

Firms often finance part of their assets with equity (ordinary, preference and retained earnings) capital, while the other part is financed by other resources such as long-term financial debts or liabilities (like bonds, bank loans and other loans) and other short-term liabilities for example trade payables (Gambacorta, Yang & Tsatsaronis, 2014). Firms can choose among many alternative financial structures. For example, short term debt financing, long term debt financing, share capital and retained earnings. A firm can also arrange lease financing, use warrants, issue convertible bonds, sign forward contracts or trade bond swaps (Ghazouani, 2013). Firms can also issue dozens of distinct securities in countless combinations to maximize overall market value (Dare & Sola, 2010). Financial structure is therefore very critical and fundamental in the business life cycle not only to maximize shareholders' wealth but also due to the impact it has on financial growth (Ishaya & Abduljeleel, 2014). The share capital of a firm is linked to the financial growth of a firm.

A joint stock firm should have capital in order to finance its activities. Funds raised by issuing shares in return for cash or other considerations. A firm usually raises its capital in the form of shares (called share capital) and debentures (debt capital) (Colombo, 2007). These funds breathe life into the business entity enabling it to commence operations thereby achieving the objectives for which it was set up. The Memorandum of Association must state the amount of capital with which the company is desired to be registered and the number of shares into which it is to be divided (Armour, 2010). It constitutes the basis of the capital structure of a company. In other words, the capital collected by a joint stock company for its business operation is known as share capital (Younus, Ishfaq, Usman & Azeem, 2014). Share capital is an indication of value contributed to a firm by its shareholders at some time in the past.

Several provisions of the Companies Act of 2006 regulate dealings with corporate share capital. A common rationalization of the share capital provisions is that they protect corporate creditors from the abuse of limited liability by shareholders (Akbarpour & Aghabeygzadeh, 2011). The idea that creditors need such protection is of course used to explain a wide range of company insolvency law doctrines (Chowdhury & Chowdhury, 2010). A company may increase its share capital either through a fresh issue of shares, or by capitalizing retained earnings with a bonus issue (Arulvel & Ajanthan, 2013). Capital may be decreased, in response to a long-term drop in the firm's net assets, by reducing the nominal value of shares through a reduction of capital pursuant to Companies Act of 2006. This requires the court's approval, but because it does not involve any direct transfer of assets to the shareholders, the creditors do not usually have any right to object.

Financial growth is a measure of efficient utilization of assets by a firm from principal business mode to generate revenues (Aburub, 2012). Financial growth is a general measure of the overall financial health of a firm over a given period (Onaolapo & Kajola, 2010). According to Buvanendra, Sridharan and Thiyagarajan (2017), the financial growth of a firm is measured as the growth in market capitalization. Market capitalization refers to the total dollar market value of a company's outstanding shares (San & Heng, 2011). Market capitalization is calculated by multiplying a form's shares outstanding by the current market price of one share (Buvanendra et al., 2017). The return on investment, return on assets, market value, and accounting profitability reflect financial growth of firms (Ongore & Kusa, 2013).

## 1.2 Statement of the Problem

Share capital in the financial structure of a firm is the total capital of a firm divided into shares. A joint stock firm should have capital in order to finance its activities. Funds raised by issuing shares in return for cash or other considerations. A firm usually raises its capital in the form of shares (called share capital) and debentures (debt capital) (Colombo, 2007). These funds breathe life into the business entity enabling it to commence operations thereby achieving the objectives for which it was set up. The Memorandum of Association must state the amount of capital with which the company is desired to be registered and the number of shares into which it is to be divided (Armour, 2010). It constitutes the basis of the financial structure of a firm. In other words, the capital collected by a joint stock company for its business operation is known as share capital (Younus *et al.,* 2014). Share capital is an indication of value contributed to a firm by its shareholders at some time in the past.

A significant number of the non-financial firms listed at the Nairobi Securities Exchange (NSE) have been experiencing declining financial performance which deters investors from investing in such firms (Muchiri, Muturi & Ngumi, 2016). The growth of non-financial firms listed at the Nairobi Securities Exchange was 3.7% in 2017 against 4.2% in 2016 (NSE, 2017). Decline in financial performance deter lenders from lending to such firms (Muchiri, Muturi & Ngumi, 2016). For instance, Kenya Airways Limited reported a net loss of Kshs. 26.2bn ($258m) for the financial year 2015-2016, up from Kshs. 25.7bn in the previous financial year (NSE, 2016). Uchumi Supermarket Limited was revived after an agreement between the Kenyan government, suppliers and debenture holders (NSE, 2017).

Ebaid (2009) investigated the impact of capital-structure choice on firm performance: empirical evidence from Egypt and found out that, share capital has a significant relationship with return on assets but insignificant relationship with return on equity. Share capital financing differ from firms to films because of the different in the financial structure. This current study focuses at non-financial firms listed at the Nairobi Securities Exchange. In a study on impact between capital structure and financial performance of Sugar companies listed at Karachi Stock Exchange Pakistan, Younus *et al.,* (2014) showed that there was weak positive correlation between share capital firm performances. Tsoy and Heshmati (2017) conducted a study on the impact of financial crises on dynamics of capital structure of listed non-financial firms in Korea and revealed that high level of leverage is associated with share capital impacting aggregate financial growth of the firm over time. However, the study did not show how share capital impacts financials growth of firms presenting conceptual gap.

A study by Oma and Memba (2018) on the effect of share capital finance on profitability of petroleum marketing firms in Kenya and share capital has a negative and insignificant effect on profitability of the firm. The study focused at petroleum firm whose mode of financial structure may differ from that of non-financial firms at the Nairobi Securities Exchange presenting contextual gap. Muchiri *et al.* (2016) examined the effects of financial structure on performance of listed Investment firms in Kenya and the findings revealed that, share capital had a significant and positive relationship with Return on Assets (ROA) and Return on Equity (ROE). The study did not explicitly indicate to what extent the share capital influence financial performance of listed Investment firms in Kenya. Further, the study focused only on Investment firms listed at the Nairobi Securities Exchange. The study intends to fill this conceptual gap by focusing on the effect of share capital on financial growth of non-financial firms listed at the Nairobi Securities Exchange.

## 1.3 Research Objective

To assess the effect of Share capital on financial growth of Non-financial firms listed at the Nairobi Securities Exchange.

## 1.4 Statistical Hypothesis

There is no significant effect of Share capital on financial growth of Non-financial firms listed at the Nairobi Securities Exchange.

## 2.0 Literature Review

This section presents the theories, empirical reviews and conceptual framework.

## 2.1 Theoretical Review

The study is guided by Market Timing Theory and Theory of Growth of the Firm.

### 2.1.1 Market Timing Theory

The market timing theory was proposed by Baker and Wurgler (2002). The market timing theory assumes that, firms time their equity issues whereby they will issue new stock when the stock price is perceived to be overvalued (high price) and repurchase their shares when there is undervaluation (low price) (Luigi & Sorin, 2009; Mostafa & Boregowda, 2014). This implies that firm's intent to take advantage of fluctuations in equity market valuations. As a result, fluctuations in stock prices will affect the firm's financial decision as well as its financial growth. The management of the firm is presumed to only issue equity when they perceive the firm's shares to be overvalued, while in case of undervaluation, the management of the firm repurchases equity and/or issue debt (Baker & Wurgler, 2002; de Bie & de Haan, 2007). The overvaluation of a firm's share results from either information asymmetry reduction or irrational investors' behavior (Danso & Adomako, 2014). Information asymmetry between management of the firm and investors reduces when the management releases information on the firm's forecasts. In case of positive forecasts, the share price rises, and overvaluation is likely to occur (Huang & Ritter, 2005).

The study by Baker and Wurgler (2002) was the first to provide evidence that equity market valuation fluctuations have a long- term effect on firm's financial structure. Subsequent studies of Leary and Roberts (2005) and Kayhan and Titman (2007), however, found only short-term effects since deviation from the target leverage ratio reversed on the long-term. As such, these authors argue that market timing is only a short-term determinant that behaves in line with the dynamic trade-off theory (Baker & Gerald, 2011). Market timing theory was relevant to the study by enabling firms decide whether to finance their investments via debt or equity. The firms either chooses to finance through equity or debt based on the market situation.

**2.1.2 Theory of Growth of the Firm**

The theory was propagated by Penrose (1959). Penrose argued that firms had no determinant to long run or optimum size, but only a constraint on current period growth rates (Penrose, 1959). According to the theory, financial means for expansion could be found through retained earnings, borrowing, and new issues of stock shares. Retained earnings are one of the most important sources to finance new projects in emerging economies where capital markets are not well developed. However, firms in the start-up period, when initial investments have not matured yet or whose investment projects are substantially larger than their current earnings, will not have enough financial means from retained earnings and will face a constraint in their growth project. Firms in this situation may seek external sources of financing; however, the extent of borrowing could be limited by internal factors like high debt-equity ratios that would expose both borrower and lender to increased risk. In other cases, financing of growth projects may be limited by shallow financial markets. Rajan and Zingales (1998) found that industrial sectors with a great need for external finance grow substantially less in countries without well-developed financial markets.

This theory is relevant to this study since it informs the dependent variable which is financial growth. The current studies which have used this theory of firm's growth are; Diaz Hermelo (2007) who conducted a study on the determinants of firm's growth: an empirical examination and Pervan and Višić (2012) who conducted a study on the Influence of firm size on its business success.

## 2.2 Empirical review

Arulvel and Ajanthan (2013) conducted a study on capital structure and financial performance of listed trading firms in Sri lanka. The study employed panel design. The results shows that, debt ratio is negatively correlated with all financial performance measures [Gross Profit (GP); Net Profit (NP); Return on Equity (ROE) and Earnings Per Share (EPS)] similarly debt-equity ratio (D/E) is negatively correlated with all financial performance measures except GP and only (D/E) ratio shows significant relationship with NP impacting aggregate financial growth of the firm over time.

Akbarpour and Aghabeygzadeh (2011) investigated the relationship between financial structure and accounting measurement for evaluating performance (ROA, ROE) for the period 2005-2010 of listed firms at Tehran Security Exchange (TSE). The study adopted panel design. The results indicated that, there was a significant relationship between financial structure and ROA, but there isn't such a significant relationship between financial structure and ROE. It was established that financial structure plays an important role in the profitability of enterprises hence financial growth.

Younus *et al.,* (2014) identified the impact between capital structure and financial performance of Sugar companies listed at Karachi Stock Exchange Pakistan (KSE Pakistan). This research includes 33 sugar companies listed in KSE Pakistan from the year of 2006-2011. Panel data research design was used. Secondary data was utilized from company's financial reports, annual reports and state bank of Pakistan in financial review for the period of six years (2006-2011). The results showed that there was weak positive correlation.

Tsoy and Heshmati (2017) conducted a study on the impact of financial crises on dynamics of capital structure of listed non-financial firms in Korea. Using a data set covering 1,159 Korean listed non-financial firms from 10 industrial sectors over the period 1985-2015, the pattern of firms' capital structure before and after the crises is investigated and the speed of adjustment towards the optimal leverage identified. The results revealed that high level of leverage is associated with share capital impacting aggregate financial growth of the firm over time. However, the study did not show how share capital impacts financials growth of firms presenting conceptual gap.

Ebaid (2009) conducted a study on the impact of capital-structure choice on firm performance: empirical evidence from Egypt. Return on Asset (ROA) and Return on Equity (ROE) were used as the measures of performance of the firm, while short term debt, long-term debt and total debt represented indicators of capital structure. Descriptive research design was adopted. The study found out that, share capital and long-term debt has a significant relationship with return on assets but insignificant relationship with return on equity. It was concluded that, capital structure changes do affect the performance of the firm impacting aggregate financial growth of the firm over time.

## 2.3 Conceptual Framework
Figure 1 shows the conceptual framework.

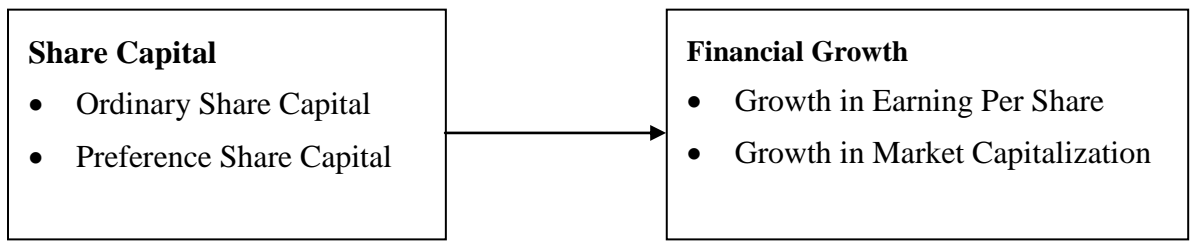

**Figure 1: Conceptual Framework**
Share capital is the independent variable. The dependent variable is financial growth measured using growth in earnings per share and growth in market capitalization.

## 3.0 Research Methodology
The study adopted positivism research philosophy. Explanatory research design was also adopted. The study conducted both descriptive statistics analysis and panel data analysis. Panel analysis permits the researcher to study the dynamics of change with short time series. The combination of time series with cross-section can enhance the quality and quantity of data in ways that would be impossible using only one of these two dimensions (Gujarati, 2009). The target population of the study comprised 45 non-financial firms listed at the NSE for a period of ten years from 1st January 2008 to 31st December 2017 (NSE, 2017). Secondary data was extracted from published audited financial statements. Panel data obtained covered a period of 10 years beginning from 2008 and ending in 2017. The panel model estimated was: -

$FG_{it} = \beta_0 + \beta_1 SC_{it} + \mu$

Where;

FG = Financial growth measured by growth in earnings per share and growth in market capitalization of firm i at time t

$\beta_0$ = Alpha coefficient representing the constant term

$\beta_i$ = Beta coefficient

SC = Share capital of firm i at time t

i = Firms listed from 2008 to 2017

t = Time period (2008-2017)

µ = Error term

## 4.0 Research Findings and Discussions

Results of the study are presented in this section. Discussion and collaboration of findings has also been conducted.

### 4.1 Descriptive Statistics

Table 1 shows the descriptive statistics for share capital, earnings per share and market capitalization.

**Table 1: Descriptive Statistics**

| Variable | Obs | Mean | Std. Dev. | Min | Max |
|---|---|---|---|---|---|
| Share Capital | 360 | 0.100219 | 0.156585 | 0.001601 | 1.139994 |
| Earnings Per Share | 360 | 6.468265 | 15.03232 | -46.744 | 100.0483 |
| Market Capitalization in million KES | 360 | 24600.00 | 77300.00 | 116.000 | 721000.00 |

Share capital had a mean of 0.100219 with a minimum of 0.001601, a maximum of 1.139994 and standard deviation of 0.156585. The mean value for earnings per share was 6.468265 with a minimum of -46.744, a maximum of 100.0483 and standard deviation of 15.03232. The mean value for market capitalization as another measure of financial growth was Kshs. 24600 million with a minimum of Kshs. 116 million and a maximum of Kshs. 721000 million.

## 4.2 Trend Analysis

This section presents the analysis of the trends of the variables. The study conducted a trend analysis to establish the movement of the variables overtime. Trend analysis for share capital, Earnings Per Share and Market Capitalization are presented in figure 2, figure 3 and figure 4 respectively.

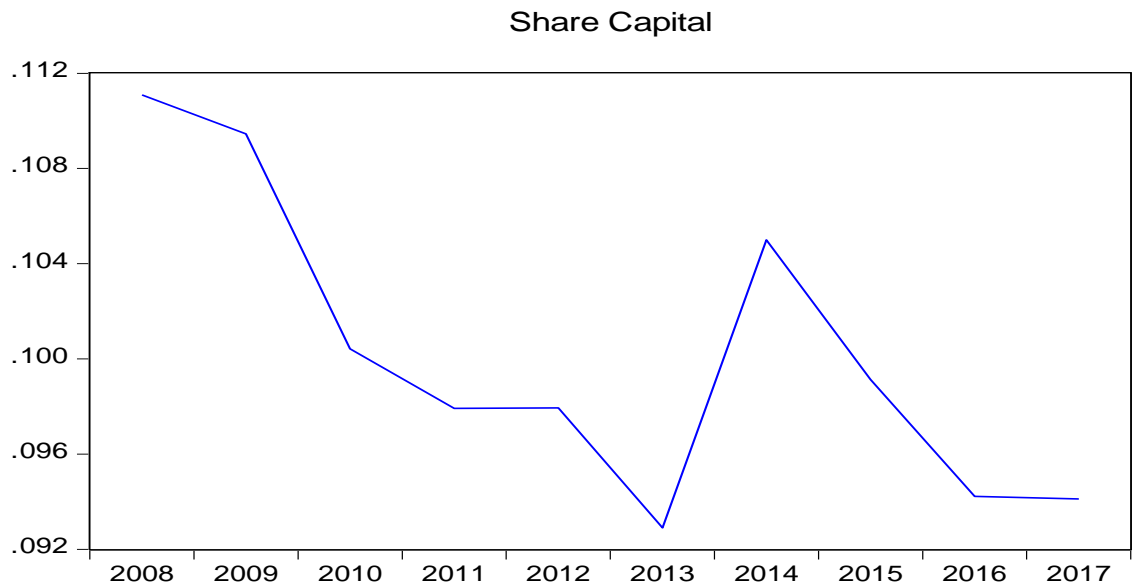

**Figure 2: Share Capital Trend Line**

Share capital was highest in 2008 and gradually declined to lowest 2013. From 2014, share capital rose gradually in 2014 before declining in the subsequent years. A joint stock company should have capital in order to finance its activities. Funds raised by issuing shares in return for cash or other considerations. According to Oma and Memba (2018) share capital has a negative but insignificant effect on profitability.

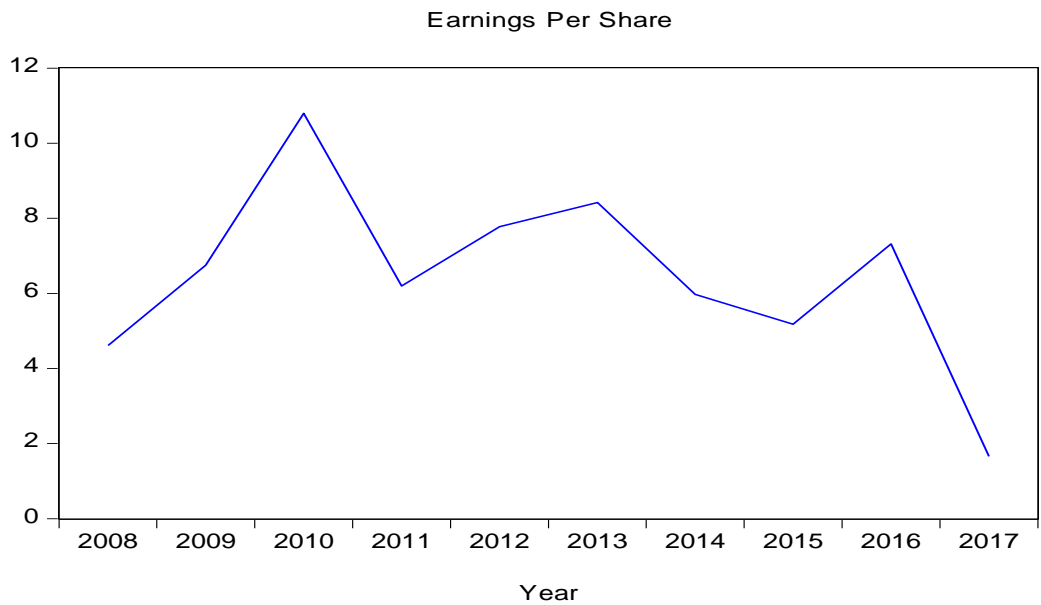

**Figure 3: Earnings per Share Trend Line**

The trend line shows that, earnings per share rose steadily from 2008 to reach highest in 2010. However, Earning Per Share dropped further from 2011 to 2012 before rising again in 2013. Earnings per share further dropped drastically to reach lowest in 2017. Earnings per share is considered as an important accounting indicator of risk, entity performance and corporate success. It is used to forecast potential growth in future share prices, because changes in earnings per share are often reflected in share price behavior. Smart and Graham (2012) concur by suggesting that an entity's growth rate is determined by performance indicators such as earnings per share which is disclosed in the financial statements of companies according to the specifications of the specific accounting standards applied in the respective country. Furthermore, authors have argued that earnings per share has become a useful investment decision tool for investors, because it indicates future prospects and growth (Mlonzi, Kruger & Ntoesane, 2011). According to Robbetze, de Villiers and Harmse (2017), earnings per share correlated best with the changing behavior of share prices.

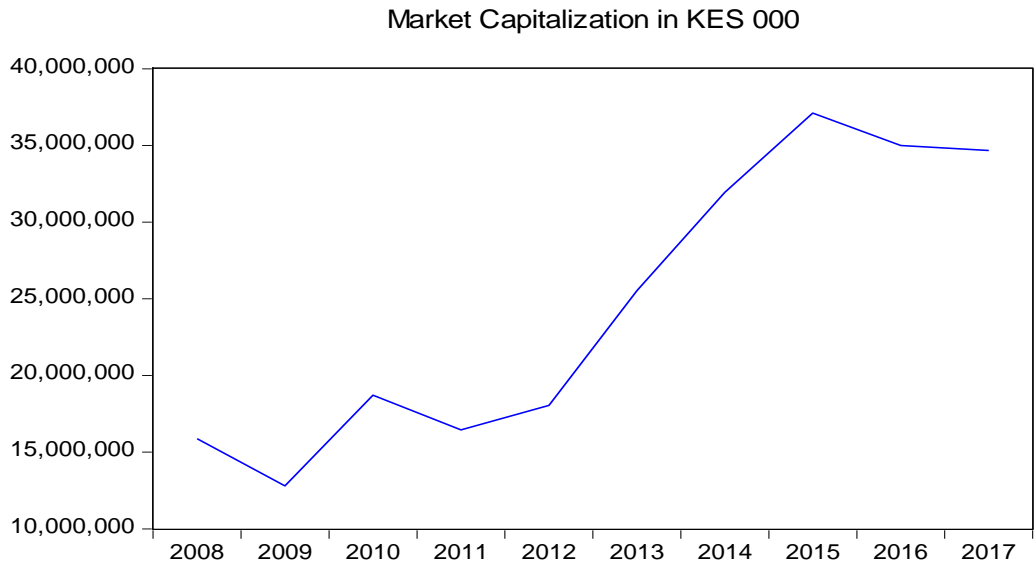

**Figure 4: Market Capitalization Trend Line**

Market capitalization as another measure of financial growth was lowest in 2009 and sharply rose to highest in 2015. Market capitalization is important in projecting the size of an organization because it shows the organization's value. Market capitalization is a measure of the value of companies and stock markets which is an on-going market valuation of a public firm whose shares are publicly traded on a stock exchange computed by multiplying the number of outstanding shares held by the shareholders with the current per share market price at a given time. A market capitalization calculation is a critical part of any stock valuation formula as it represents the total market value of all the company's outstanding shares. This represents the value the market has placed on the value of a company's equity. As outstanding stock is bought and sold in public markets, market capitalization could be used as a proxy for the public opinion of a company's net worth and is a determining factor in some forms of stock valuation. Market capitalization represents the public consensus on the value of a company's equity. According to Koila, Kiru and Koima (2018) using random effects model also revealed that market capitalization cannot be used to predict the outcome of return on equity within the listed firms at the Nairobi Securities Exchange.

### 4.3 Correlation Analysis

Table 2 shows the correlation matrix of share capital and growth in earnings per share as a measure of financial growth.

**Table 2: Correlation between Share Capital and Growth in Earnings Per Share (EPS)**

|  | Growth in EPS | Share Capital |
|---|---|---|
| Growth in EPS | 1.000 |  |
| Share Capital | 0.150 | 1.000 |

It was also established that share capital and growth in earnings per share are positively associated. Share capital is an indication of value contributed to a firm by its shareholders at some time in the past. A joint stock company should have capital in order to finance its activities. Funds raised by issuing shares in return for cash or other considerations. A firm usually raises its capital in the form of shares. The results contrast Oma and Memba (2018) that share capital has a negative and insignificant effect on profitability. Table 3 shows the correlation matrix of share capital and growth in market capitalization as a measure of financial growth.

**Table 3: Correlation between Share Capital and Growth in Market Capitalization**

|  | Growth in Market Capitalization | Share Capital |
|---|---|---|
| Growth in Market Capitalization | 1.000 |  |
| Share Capital | 0.155 | 1.000 |

The correlation established that share capital and growth in market capitalization are positively associated. Share capital is an indication of value contributed to a firm by its shareholders at some time in the past. A joint stock company should have capital in order to finance its activities. A firm usually raises its capital in the form of shares. According to Ebaid (2009) that share capital has significant relationship with return on assets but not with return on equity.

**4.4 Panel Regression Analysis**

**4.4.1 Effect of Share Capital on Financial Growth**

Panel regression analysis was conducted on share capital and financial growth measured using growth in earnings per share and growth in market capitalization. Random model was estimated to determine whether there was a significant relationship between share capital and growth in earnings per share. Table 4 presents the panel regression model on share capital and growth in earnings per share as a measure of financial growth.

**Table 4: Effect of Share Capital on Growth in Earning Per Share (EPS)**

| Growth in EPS | Coef. | Std. Err. | T | P>|t| | [95% Conf. | Interval] |
|---|---|---|---|---|---|---|
| Share Capital | 0.026871 | 0.009395 | 2.86 | 0.004 | 0.008456 | 0.045285 |
| _cons | 1.547499 | 1.169374 | 1.32 | 0.186 | -0.74443 | 3.83943 |
| R-squared: | = 0.3273 | | | | | |
| Wald chi2(1) | = 8.18 | | | | | |
| Prob > chi2 | = 0.0042 | | | | | |

The fitted model from the result is

Growth in EPS = 1.547499 + 0.026871SC

Where: EPS = Earnings Per Share

SC = Share Capital

As presented in Table 4, the coefficient of determination R-Square is 0.3273. The model indicates that, share capital explains 32.73% of variation in the growth in earnings per share as a measure of financial growth. This means 32.73% of variation in the growth in earnings per share as a measure of financial growth is influenced by share capital. The findings further confirm that, the relationship between share capital and financial growth as measured by growth in earnings per share is positive and significant with a coefficient of (β=0.026871, p=0.004). This implies that, there exist a positive and significant relationship between share capital and financial growth as measured by growth in earnings per share since the beta coefficient value was positive and the p-value 0.004< 0.05. This means that, a unitary increase in share capital leads to financial growth as measured by growth in earnings per share by 0.026871 units holding other factors of financial structure constant. An entity that trades equity instruments on public markets is required to disclose earnings per share as part of its financial statements. Earnings per share is a calculation that allocates a firm's profits to each of its ordinary shares. It serves as an indication of profitability by measuring the entity's performance in relation to share capital that is employed to generate such returns. According to Robbetze, de Villiers and Harmse (2017) earnings per share correlates best with the changing behavior of share prices. Table 5 presents the panel regression model on share capital and growth in market capitalization as a measure of financial growth.

**Table 5: Effect of Share Capital and Growth in Market Capitalization**

| Growth in Market Capitalization | Coef. | Std. Err. | T | P>\|t\| | [95% Conf. Interval] | |
|---|---|---|---|---|---|---|
| Share Capital | 0.027239 | 0.009855 | 2.76 | 0.006 | 0.007922 | 0.046555 |
| _cons | 4.985739 | 1.386054 | 3.6 | 0.000 | 2.269124 | 7.702355 |
| R-squared: | = 0.1162 | | | | | |
| Wald chi2(1) | = 7.64 | | | | | |
| Prob > chi2 | = 0.0057 | | | | | |

The fitted model from the result is

Growth in market capitalization = 4.985739 + 0.027239SC

Where: SC = Share capital

As presented in the Table 5, the coefficient of determination R-Square is 0.1162. The model indicates that, the share capital explains 11.62% of variations in the growth in market capitalization as a measure of financial growth. This means 11.62% of variation in the growth in market capitalization as a measure of financial growth is influenced by share capital. The findings further confirm that, the panel regression model for share capital and growth in market capitalization as a measure of financial growth is positive and significant with a coefficient of ($\beta$=0.027239, p=0.0057). This implies that, there exist a positive and significant relationship between share capital and financial growth as measured by growth in market capitalization since the coefficient value was positive and the p-value 0.0057<0.05. This means that, a unitary increase in share capital leads to growth in market capitalization as a measure of financial growth by 0.027239 units holding other elements of financial structure constant. Share capital is considered a more accurate estimate of a firm's actual net worth. It is all funds raised by the firm in exchange for shares of either a common or preferred stock. Share capital is also referred to as equity financing. A firm can increase its share capital by obtaining authorization to issue and sell additional shares. Share capital consists of all funds raised by a firm in exchange for shares of either common or preferred shares of stock. The amount of share capital or equity financing a company has can change over time. A firm that wishes to raise more equity can obtain authorization to issue and sell additional shares, thereby increasing its share capital. The results are in line with Ebaid (2009) share capital has positive and significant relationship with market capitalization.

### 4.5 Hypothesis Testing

The hypothesis was tested using p-value method. The acceptance/rejection criterion was that, if the p-value is greater than the significance level of 0.05, we fail to reject the Ho but if it's less than 0.05 level of significance, the Ho is rejected. The null hypothesis was that, there is no significant effect of share capital on financial growth of Non-financial firms listed at the Nairobi Securities Exchange. The results in Table 4 shows that, share capital and growth in earning per share are positively and significantly related with p-value =0.004<0.05. Further, the results in Table 5 shows that share capital and growth in market capitalization is positively and significantly related with p-value =0.006<0.05. The null hypothesis was therefore rejected and concluded that there is a significant effect of share capital on financial growth of Non-financial firms listed at the Nairobi Securities Exchange.

### 5.0 Conclusions

The study concluded that, share capital has a positive and significant relationship with financial growth measured using growth in earning per share. A joint stock company should have capital in order to finance its activities. Share capital is an indication of value contributed to a firm by its shareholders at some time in the past. It was further concluded that, share capital has a positive and significant relationship with financial growth measured using growth in market capitalization.

Share capital (shareholders' capital, equity capital, contributed capital, or paid-in capital) is the amount invested by a firm's shareholders for use in the business. When a firm is first created, if its only asset is the cash invested by the shareholders, the balance sheet is balanced with cash on the left and share capital on the right side. Share capital is a major line item but is sometimes broken out by firms into the different types of equity issued. There can be common stock and preferred stock, which are reported at their par value or face value. A firm may increase its share capital either through a fresh issue of shares, or by capitalizing retained earnings with a bonus issue.

**6.0 Recommendations**

Share capital is the funds that a firm raises through the issuance of common or preferred stock. With additional public offerings, the amount of share capital or equity financing that a firm has can change over time. The study revealed that, share capital significantly influences financial growth. The study recommends for the Non- financial firms to utilize equity financing as a way of raising capital for major expansions, asset growth or acquisitions which may require heavy funding. In this way, firms will be assured of improved performance as well as high financial growth. The study also recommends for substantial firm financing through equity. Equity financing is important to any firm, if the proceeds are used to invest in projects which eventually bring growth to the firm.

Non-financial firms may consider using share capital as mode of fast and easily available form of financing. Raising equity via share sales is also very flexible. The business has full control over how many shares to issue, what to initially charge for them and when it wishes to issue them. It can also issue further shares in the future if it wishes to raise more money. The firm can also decide on the type of shares it issues and what rights these give the shareholders, and it can also repurchase issued shares if desired.

Ishaya, L. C., & Abduljeleel, B. O. (2014). Capital Structure and Profitability of Nigerian Quoted: The Agency Cost Theory Perspective. *American International Journal of Social Science, 3*(1), *139-140.*

Kayhan, A., & Titman, S. (2007). Firms' histories and their capital structures. *Journal of financial Economics, 83(1), 1-32.*

Koila, E., Kiru, K., & Koima, J., K. (2018). Relationship between Share Market Capitalization and Performance of Listed Firms at Nairobi Securities Exchange Limited, Kenya Relationship between Share Market, *Scholars Journal of Economics, Business and Management, 5(3), 1-23.*

Leary, M. T., & Roberts, M. R. (2005). Do firms rebalance their capital structures? *The journal of finance, 60(6), 2575-2619.*

Luigi, P., & Sorin, V. (2009). A review of the capital structure theories. *Annals of Faculty of Economics, 3(1), 315-320.*

Menike M. G. P. & U. S. Prabath (2014). The Impact of Accounting Variables on Stock Price: Evidence from the Colombo Stock Exchange, Sri Lanka: *International Journal of Business and Management*; 9 (5), 210–221.

Mlonzi, V.F., Kruger, J. & Ntoesane, M.G. (2011). Share price reaction to earnings announcement on the JSE-AltX: A test for mark efficiency. *Southern African Business review*, 15(3), 142-166.

Modigliani, F. & Miller, M., (1958). The cost of capital, corporation finance and the theory of investment. *The American Economic Review*, 48 (3), 261 - 297.

Modigliani, F. and Miller, M. (1963). Corporate income taxes and the cost of capital: a correction. *American Economic Review*, 53 (3), 443-453.

Mostafa, H. T., & Boregowda, S. (2014). A brief review of capital structure theories. *Research Journal of Recent Sciences, ISSN, 2277, 2502.*

Muchiri, M. J., Muturi, W. M., & Ngumi, P. M. (2016). Relationship between Financial Structure and Financial Performance of Firms Listed at East Africa Securities Exchanges. *Journal of Emerging Issues in Economics, Finance and Banking,* 7(5), 1734-1755.

Oma, M., D & Memba, F., S. (2018). The Effect of Share Capital Finance on Profitability of Petroleum Marketing Firms in Kenya, *International Journal of Economics, Commerce and Management,6(1),410-422.*
43